\documentclass[fleqn,usenatbib]{mnras}

\usepackage[T1]{fontenc}

\DeclareRobustCommand{\VAN}[3]{#2}
\let\VANthebibliography\thebibliography
\def\thebibliography{\DeclareRobustCommand{\VAN}[3]{##3}\VANthebibliography}

\newcommand\be{\begin{equation}}
\newcommand\ee{\end{equation}}
\newcommand\bea{\begin{eqnarray}}
\newcommand\eea{\end{eqnarray}}

\usepackage{epsfig}
\usepackage{amssymb}
\usepackage{gensymb}

\usepackage{color}
\definecolor{red}{rgb}{0.7,0,0}
\definecolor{blue}{rgb}{0,0,0.7}
\definecolor{purple}{rgb}{0.7,0,0.7}

\def\refe#1{{}{{#1}}}
\def\refee#1{{}{{#1}}}
\def\refeee#1{{}{{#1}}}

\title[BBHs' Observables vs SBH]
{\refee{Can we disentangle between the emission of an accretion disc around a single black hole and a circumbinary disc ?}}

\author[Varniere et al.]{
Peggy Varniere,$^{1,2}$\thanks{E-mail: varniere@apc.in2p3.fr}
Rapha\"{e}l Mignon-Risse,$^{3,4}$\thanks{E-mail: raphael.mignon-risse@ntnu.no}
Fabien Casse$^{1}$
\\
$^{1}$ Universit\'e Paris Cit\'e, CNRS, Astroparticule et Cosmologie, F-75013 Paris, France\\
$^{2}$ Universit\'e Paris-Saclay, Universit\'e Paris Cit\'e,CEA, CNRS, AIM, 91191, Gif-sur-Yvette, France \\ 
$^{3}$ Universit\'e Paris Cit\'e , CNRS, CNES, Astroparticule et Cosmologie, F-75013 Paris, France	\\
$^{4}$ Department of Physics, Norwegian University of Science and Technology, NO-7491 Trondheim, Norway\\
}

\date{Received  / Accepted }

\begin{document}
\label{firstpage}
\pagerange{\pageref{firstpage}--\pageref{lastpage}}
\maketitle

\begin{abstract}
{The detection of gravitational waves from binary black holes (BBHs) started the hunt for their pre-merger electromagnetic emission. In that respect, numerical simulations have been looking for the  \lq smoking gun{\rq} signal that could help identify pre-merger systems.}
{Here we study if any of the expected features of circumbinary discs, such as the periodic modulation from the orbiting \lq lump{\rq}, could be used to identify pre-merger BBHs or if they could be easily \refe{confused with} other systems. Indeed, while the timing feature associated with the \lq lump{\rq} seems to be present for a large part of the parameter space defined by the binary separation and mass ratio \refe{in circular binaries},  it was recently proposed to form thanks to an instability  occurring naturally at the edge of accretion discs around single black holes (SBH).  
\refe{In order to check if features of a circumbinary disc could be reproduced by a SBH system, we search for at least one SBH fit able to replicate
the given synthetic observations of a circumbinary disc.}}
{We found that many of the features from a circumbinary disc can be reproduced by a SBH system with different masses, distances or inner disc positions. 
}
{Interestingly, while we can always find a SBH model providing a good enough fit to the data, the presence of two variabilities, associated with the lump and the binary\refe{, or binary-lump beat,} period, is a necessary condition for
   a wide range of \refe{BBH} system parameters and should be used as a test to disqualify some BBH candidates.
 }
 \end{abstract}


\section{Introduction}

The detection of gravitational waves (GW) from the merger of compact objects and their electromagnetic counterparts has opened a new era of 
multi-messenger astronomy both in observations \citep{abbott_gravitational_2017, graham_candidate_2020} but also in numerical simulations 
as they are currently the only way to constrain what  observables would help us detect a  
BBH system before or during the early stage of gravitational wave detection.
\refe{The major part of the numerical studies of pre-merger binary black-holes (BBHs) \refee{is} looking at the system between the moment a circumbinary disc is formed down to the merger 
\cite[e.g. reviews by][]{gold_relativistic_2019,dorazio_observational_2023}.} 
Those  numerical studies converged toward a system \refe{where the circumbinary disc is feeding the individual black holes through two streams which}  
might form \refe{some} intermittent disc-like structures around each black hole \citep{macfadyen_eccentric_2008,shi_three-dimensional_2012,noble_circumbinary_2012, 
dorazio_accretion_2013,gold_accretion_2014,shi_three-dimensional_2015,armengol_circumbinary_2021,tiede_how_2021,liu_evolution_2021}. 
One of the most interesting features is the presence of a strong  $m=1$ azimuthal mode 
at the inner edge of the circumbinary disc leading to an overdensity that was dubbed \lq lump\rq\ by \cite{shi_three-dimensional_2012}. Such an overdense lump  is promising as it could lead to 
an observable feature in the form a 
 modulation of the electromagnetic flux, especially in the case of a high inclination system
 \citep{noble_circumbinary_2012,dorazio_accretion_2013,farris_binary_2014,noble_mass-ratio_2021-1}, in turn  this signal could allow us to identify pre-merger systems \refe{with a circumbinary disc}. 
  \refe{It is worth noting that this overall structure is also seen in  Newtonian simulations  \cite[see the extensive review on circumbinary disc by][]{LaiMunoz23} and across codes \cite[as  is  shown in the Santa Barbara Binary-Disc code comparison in][]{codecomp_SB}, hence making it a robust feature of circumbinary discs \refe{around circular binaries}. }
 
 In  \cite{MVC23b}  we proposed\refe{ an instability \citep[the Rossby-Wave Instability, RWI, ][]{Lovelace78,Lovelace99,Li00} } 
 to be at the origin of the $m=1$ mode that gives rise to the lump at the edge of the circumbinary disc which 
 would create a quasi-periodic modulation of the lightcurve (e.g. \citealt{tang_late_2018}).
This was a beneficial step as it gives us a better understanding of what causes the lump, hence we can \refe{better predict} in which systems it would occur, which in turn tells us \refe{that}
such observables are expected for a large range of mass-ratio and time-to-merger. 
At the same time,  it also raises the issue of how to distinguish those \refe{observables from  the ones arising }
\refe{from the accretion disc orbiting {around}} 
a single black hole \citep[see for example][]{V20}. 
Indeed, if \refe{one wants to use  the \refe{lump-}
induced variability as a way to identify pre-merger BBHs, we 
need to make sure that no other systems, namely a {SBH} system, could exhibit the same observational features.} 
\refee{If the observable cannot be untangled, it could not only mean that some BBHs have been misclassified as single black holes,
but also that some of our candidate BBHs are in fact single black-hole systems with quasi-periodic variability. Hence, a more thorough examination is required.}
\newline
  
 \refee{To that effect, we will first briefly present}
    which stage of the pre-merger BBH system we are focusing on and what are the \refee{potential} observables that are associated with it. 
   We will also briefly present the simulations from which \refee{our synthetic observations} 
   were computed. 
   \refe{As we are looking for a possible match between observables from circumbinary discs and SBH accretion discs, the actual origin of the lump is not relevant in this study but rather its impact on the observables. As such we will not discuss in detail these simulations but only display the basic setup from which they originate.}

 \noindent In the following two sections we will focus on specific circumbinary observables and their potential matches from SBH systems. 
 \refee{Our approach will be to take each observable and check if the SBH model can be disproved or shown to lack some ingredients hinting at a more complex system.}
 The first one  will be the full spectral energy distribution (SED) of the circumbinary disc that will be confronted with a few SBH systems able to fully reproduce its observed SED. 
  Then  we will focus on 
   the temporal variability of circumbinary discs and compared it with SBH systems that could match it.
  \refee{Here the aim is to see how easy it would be simply fit the BBH's observables with a SBH.}
   Lastly, we will look at what combination of observables would be needed in order to \refee{distinguish} 
   both systems. 
   In turn this will allow us to check when the lump's impact on the observables can be confirmed as a key component to identify pre-merger systems.

 \section{\refe{Binary black hole versus single black-hole system}}
    
    \refe{For both systems we are using the units where the gravitational constant $\mathrm{G}$ and velocity of the light $\mathrm{c}$ are set to unity $\mathrm{G} \,{=}\, \mathrm{c} \,{=}1$ and the gravitational radius $r_g= \mathrm{G} M/ \mathrm{c}^2$ is computed with the total mass $M$ of the system while the mass ratio between masses in a BBH system is labelled by $q$ (with $q\leq1$).}

    
 \subsection{\refe{Explored stage of the pre-merger BBH system}}    
 \begin{figure}
\includegraphics[width=9cm]{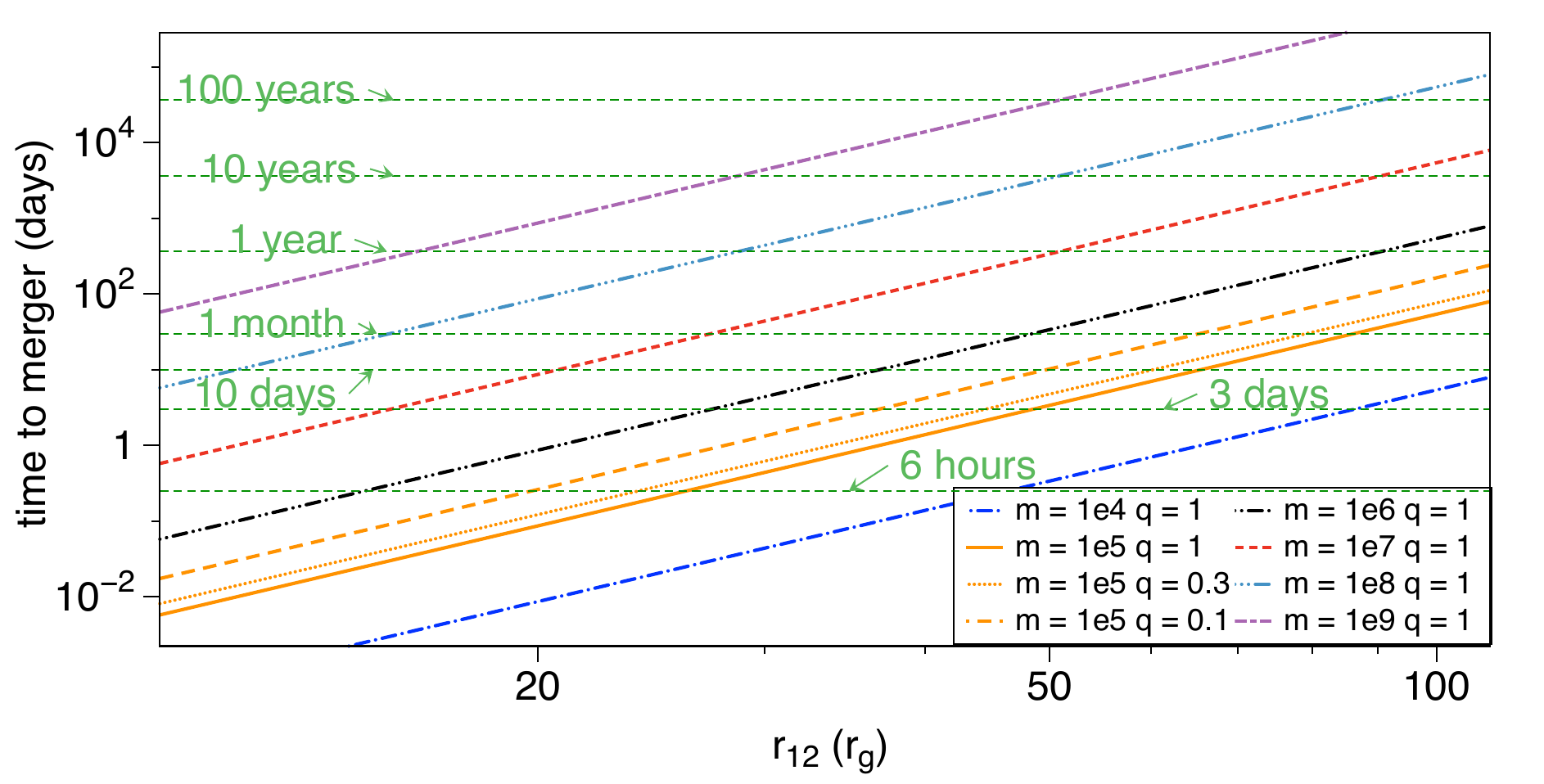}
\caption{\refe{Time to merger for BBHs as a function of the binary separation (in units of $\mathrm{r_g}$) for several binary total masses \refee{(in solar masses)} and mass ratios.}}
\label{fig:tmerger}  
\end{figure}
     \refe{In this paper we choose to focus on the pre-merger  stage where the BBHs are surrounded by a circumbinary disc while being separated by up to a few tens of gravitational radii,
      prior to binary-circumbinary disc decoupling \citep{armitage_accretion_2002}\footnote{\refee{The separation at decoupling depends on various parameters and processes, e.g., the ability of the circumbinary disc to transport angular momentum, and can occur at very small separation in some cases \citep{noble_circumbinary_2012}. Here, by assuming circular orbits we focus on the pre-decoupling stage}.}. 
     A consensus has arisen on this {pre-decoupling} stage 
     where it is believed that it       
     exhibits a promising variability coming from the \lq lump\rq\ which could be used to distinguish binary systems from SBHs \citep[][]{macfadyen_eccentric_2008,shi_three-dimensional_2012,noble_circumbinary_2012, 
dorazio_accretion_2013,gold_accretion_2014,shi_three-dimensional_2015,armengol_circumbinary_2021,tiede_how_2021,liu_evolution_2021}}. 
     
      \noindent\refe{The existence of a circumbinary disc around the BBH imposes an upper limit upon the binary separation between the two black holes as, at very large separation, 
      the original individual accretion discs are not yet rearranged into a circumbinary disc by the tidal forces of the binary. 
      While there is no firm consensus on the separation below which a circumbinary disc is {likely} 
      to exist, we have decided to restrain ourselves  to separations no larger than  $100 \, \mathrm{r_g}$. 
      Setting such maximal separation leads to  binary timescales that provide interesting estimates of the time period during which such systems can potentially be detected.}
    \noindent\refe{Considering such maximum separation,}\refe{  Fig.~\ref{fig:tmerger} displays the upper limit of the timespan 
    corresponding to our assumption, under the condition of the existence of a circumbinary disc, computed using the lowest-order post-Newtonian formula  \citep{peters_gravitational_1964}. 
      }\noindent \refe{We see that systems with total mass $M{>}10^5 \mathrm{M_\odot}$ will spend a long enough time in this pre-merger stage to reach detectability for current and future high-energy observatories \citep{mangiagli_massive_2022}, 
      especially if the system has been flagged as a 
        BBH candidate prior to forming a circumbinary disc \citep[see for example][for a study of pre-circumbinary disc system]{VCDA24}. For that reason we will focus the rest of the
        paper  on those supermassive BBH systems, even though the results presented here are independent of the system's mass.
  }
  
 \subsection{\refe{Origin of the synthetic observations used in the paper}}
 
 


     \refe{The main objective of the present paper is to confront observables characterising accretion discs orbiting a SBH to observables characterising circumbinary discs. 
     In order to do so it is {desirable} 
     to have synthetic observations based on GR fluid simulations, as analytical models \citep{roedig_observational_2014} 
    are unable to describe the complex flow morphology we described above and Newtonian fluid description failing in the relativistic regime. 
     In \refe{our} previous works, such synthetic observations were already performed with the exception of the light-curve for the SBH case \refee{which was not}
        produced for a supermassive 
     BH system in the original publication. 
     All of the fluid simulations and synthetic observations were produced by {\tt NOVAs}, our Numerical Observatory of Violent Accreting systems \citep{VCH18XMM} for the SBH system, or by {\tt e-NOVAs}, its extension to {BBHs} \citep{MVC23a}.  
     As a result we only succinctly present the code below and orient the reader to the original publications 
     of the different simulations for more details.}



%
%
%
%
  
  \noindent \refe{Fluid simulations were performed in a 2D framework with {\tt GRAMRVAC} \citep{CV18,MVC23a}, a general-relativistic hydrodynamical code using either a Kerr metric for SBH systems or a newly implemented approximated BBH metric \citep{ireland16} for circumbinary disc simulations. Synthetic observations associated with the aforementioned fluid simulations were obtained by the use of a ray-tracing code}  {\tt gyoto} \citep{Vin11}, in which we implemented \refe{the very same metrics as in the GR fluid simulations.} 
  {\tt gyoto} ray-traces the emissions back from the observer toward the disc where it uses the temperature map produced by the GR simulation to compute the 
       disc blackbody emission.
\refe{All the synthetic observations of circumbinary discs used in this paper were obtained using the method of \cite{MVC23a}, with
 the fluid simulations presented in \cite{MVC23b}.
        For the SBH case, the synthetic observations {and the fluid simulations \citep{V20}} were performed using the simpler Kerr metric. 
        Both types of synthetic observations were obtained using 
        the same procedure as presented in \cite{V20}, namely calculating the synthetic emission without 
        the fast-light approximation.\\
   It is noteworthy that 
   the inner edge location of the surrounding disc depends on the nature of the central object \citep[near the last stable orbit for SBH 
   systems and approximately twice the orbital separation in BBH system,][]{artymowicz_dynamics_1994}. 
  } The aforementioned GR fluid simulations initial setup was with gas everywhere close to gravito-centrifugal equilibrium 
    and we let the gravity carve out the inner edge of each disc
   while having similar mass distribution.
   In those simulations, an inner edge of the disc forms \refe{near its expected radius} 
\refee{and was able to form a lump for the explored mass ratio $q\in [0.1,1]$\citep{MVC23b}.
At the separation of twenty gravitational radii, this gave a lump period of about six binary periods.}
   When producing the lightcurves\refee{, and spectral energy distribution,} of each systems we only use snapshots  after the instability has reached saturation until the end of the 
   simulation\footnote{While we stopped the runs due to computational time limits, we still have up to a hundred orbits of the inner edge of the disc
  in some of the runs.}
   \refee{, {\em i.e.}}  
    tens to hundreds of inner edge orbits.  
   The lightcurves \refee{are} computed over \refee{the entire blackbody spectrum} 
     so that  different masses will have the same lightcurve but with their
     maximum emissivity pointing in a different band (see Fig.~\ref{fig:spectrum_heavy} and Fig.~\ref{fig:spectrum_trunc}). 
    We then followed the dynamical evolution of the discs leading to time-varying synthetic observations. 
    The time-variability in both types of simulation is induced by the 
   presence of a fluid instability (e.g. the RWI) leading to a non-axisymmetric pattern in the disc that directly affects the luminosity of the respective discs.

    \subsection{\refe{Observables of BBH and SBH systems} }   
    \refe{According to previous works focusing on the dynamics of the circumbinary discs in BBH systems
     \refeee{\citep[in a wide variety of gravity, thermodynamics and excision setups see for example][]{macfadyen_eccentric_2008,shi_three-dimensional_2012,noble_circumbinary_2012, 
dorazio_accretion_2013,gold_accretion_2014,shi_three-dimensional_2015,armengol_circumbinary_2021,tiede_how_2021,liu_evolution_2021,codecomp_SB}},
     two strong observables can be found,
     namely 
     \begin{itemize}
     \item the thermal  {spectrum from the} circumbinary disc whose inner edge lies around twice the binary separation \refeee{leading to a peak in the UV/optical for binary masses
     in the range $M=10^{4-10}\mathrm{M_\odot}$
     \citep{roedig_observational_2014,farris_characteristic_2015,dascoli_electromagnetic_2018,gutierrez_electromagnetic_2022,krauth_disappearing_2023,franchini_emission_2024}}\\    
     \item  a dominant flux modulation, linked with the presence of the lump at the inner edge of the circumbinary disc,  and related to the Keplerian frequency 
     around the position of the lump. \refee{A smaller modulation, related to the much shorter binary period (owing to the spiral arms or to the binary-lump beat), is also present}
     \refeee{\citep{tang_late_2018,dascoli_electromagnetic_2018,duffell_circumbinary_2020,WesternacherSchneider22,franchini_emission_2024,2024Cocchiararo}.}
     \end{itemize}} 
     
\noindent  \refe{In order to reproduce the first observable in the SBH case, full GR simulations are not required as simple observational models like 
    {\tt diskbb\footnote{spectrum from an accretion disc consisting of multiple blackbody components  \url{https://heasarc.gsfc.nasa.gov/xanadu/xspec/manual/node166.html}}} or 
    {\tt kerrbb\footnote{multi-temperature blackbody model for a thin, steady state, general relativistic accretion disc around a Kerr BH
     \url{https://heasarc.gsfc.nasa.gov/xanadu/xspec/manual/node189.html  }}}   are often used for fitting accretion discs around SBHs.}
    \refe{Only when we consider the timing aspect of SBH systems, \refee{we do} turn toward GR numerical simulations as we are still lacking observational model for 
 Quasi-Periodic Oscillations (QPO)  beyond a sinusoidal fit. 
In the post-processing step, emission and BH parameters are chosen to be consistent with a supermassive BH disc.}
 \\  

\section{\refe{Reproducing } a circumbinary disc spectrum with a single black hole system}
   One of the first observables we are able to obtain for any {supermassive} \refe{BH} system is its SED, 
   generally comprised of several non-continuous 
     bands that cover the overall behaviour of the source.
 It is therefore interesting to explore how different the SED of a circumbinary disc will be from a disc around a single \refe{BH}. 
   \refee{The presence of the lump and spiral structure in the circumbinary disc causes a relatively strong modulation of the flux, 
hence emphasizing that a non-negligible part of the flux is present within those structures. It is therefore not entirely straightforward that 
a featureless monotonic disc could reproduce the SED of a circumbinary disc.}
    We are particularly interested to see \refee{under which conditions one}  could match a circumbinary SED to a SBH one, basically looking at the possibility that we have already observed a BBH  
     system which was \lq misclassified\rq\  \refee{as a single {BH} if solely basing the comparison upon the SED of the system}
     \footnote{\refee{In theory, this also means that some BBH candidates could actually be a single black-hole systems. But, as BBH candidates are more often identified through their 
     variability than their SED, such misclassification is less likely to happen when looking at the SED fitting. }
    }.\\
    
\refee{We choose to focus only on the emission from the disc versus circumbinary disc as we want to see if they are distinguishable. 
     In a more realistic system  contaminants would be added to both emissions which would complexify the fit, but by showing how well we can fit the disc spectrum, any added structure
     on top of them would only amplify our inability to distinguish between the two}.     
    \refeee{Among those potentially added structures there are the individual accretion structures around each BH which are not accessible through the simulation of \cite{MVC23b} in which the
    inner region is excised. 
   Nevertheless, it was shown in the Santa Barbara Binary-Disk code comparison \citep{codecomp_SB}, that having an excision region does not impact the existence of 
    either  lump or spiral arms  in the simulation.  In the presence of individual accretion structures and streams, those would be at higher energy
    than the CBD \cite[see for example the representation by][]{roedig_observational_2014,2024Cocchiararo} and both components, CBD and individual structures, would be separated by a \lq notch\rq\ or 
    at least a lower emission\footnote{compared to a full accretion disc without cavity.} in the full SED. \\
\noindent    Hence, we chose to focus on the circumbinary disc spectrum which has the advantage of staying present, though decoupled from the binary evolution in later stage, even after individual accretion structure become unstable 
    as the binary separation shrinks \citep{krauth_disappearing_2023}.}

\subsection{Evolution of the circumbinary disc edge}

   \noindent  As the system we are looking to identify is comprised of the central object, either a binary or single \refe{BH}, and its associated disc, it is important to first characterize the disc position in 
       our {BBH} simulations 
       as it will directly impact the energy spectrum.     
    We showed  in \cite{MVC23b} that the inner region of the circumbinary disc has a fast evolution which might be hard to reproduce with a SBH disc.
     It is therefore interesting, even before \refe{looking at} 
     the circumbinary disc emission, to look at the instantaneous evolution of the outer edge of the cavity carved by 
     the {BBH} which, in turn, determines the position of the inner edge of the disc and hence its overall emission. \\
   Hitherto, as can be seen \refe{in all numerical simulations of circumbinary BBH discs (see e.g. \citealt{codecomp_SB} and references therein),}
   the inner region of the circumbinary disc is not circular. In order to get an estimate of the position of the inner edge of the disc we first average azimuthally the density and 
   then follow the position of the outer edge of the cavity \refe{defined using a density threshold}. 
  \begin{figure}
\includegraphics[width=8.5cm]{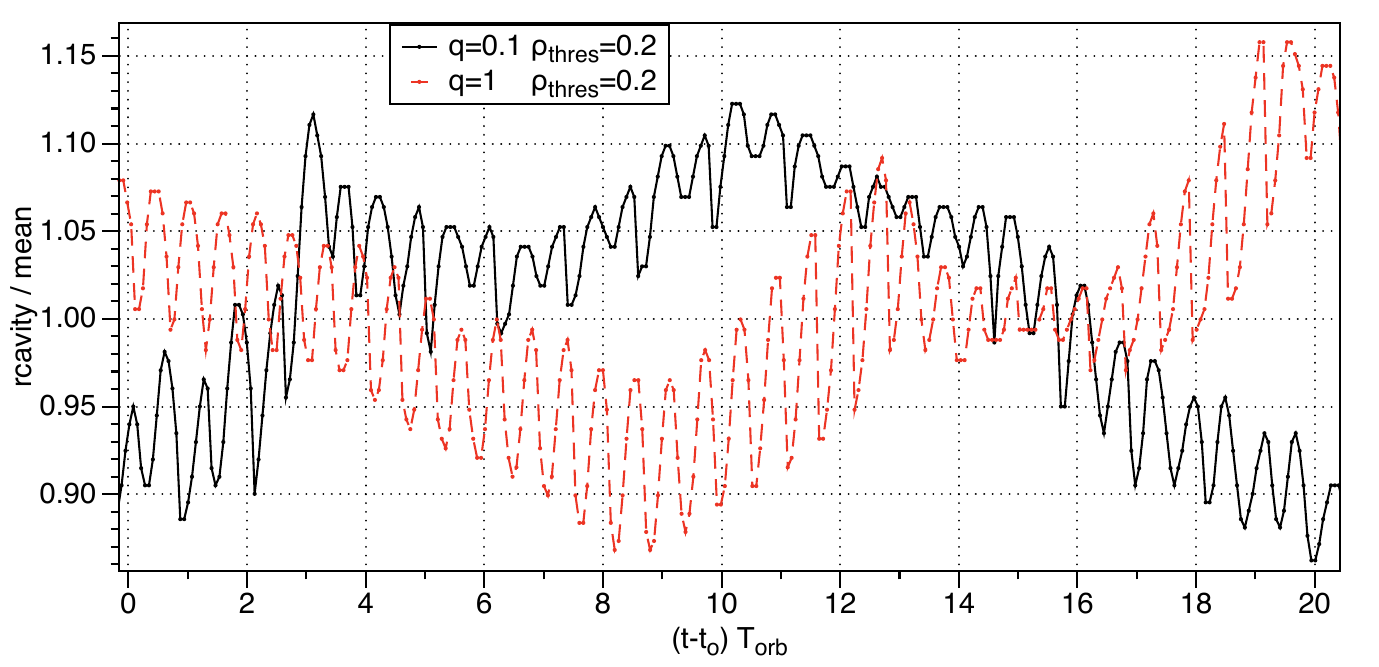}
\caption{\refe{Temporal evolution of the ratio of the outer radius of the cavity carved by the BBH divided by its mean value, for the equal-mass $q=1$ case (dash-red line) and for the $q=0.1$ (black line) case. In both cases the binary separation is $20\, \mathrm{r_g}$  \refee{with a lump period of about six binary periods,} and both simulations are shown at a similar stage where the \lq lump\rq\  is well established 
\refee{(with $t_o^{q=0.1}=144 \ T_\mathrm{orb}$ and $t_o^{q=1}=110\  T_\mathrm{orb}$ respectively to show both curves
on the same axis). }}  
}
\label{fig:rcavity}
\end{figure}
      Fig.~\ref{fig:rcavity}  shows \refe{this radius} 
      in the latter stage of circumbinary disc 
      simulations
      , meaning  when the   
      lump \refe{is fully developed and} has a strong impact on the edge of the disc \refee{(with $t_o^{q=0.1}=144 \ T_\mathrm{orb}$ and $t_o^{q=1}=110\  T_\mathrm{orb}$ respectively)}.
       \refe{In these 
      simulations, the separation is 
      $20\ r_g$ 
      while the inner edge of the disc is initially located  at a radius smaller than two times the separation. As  simulations go on, material below two times the separation is rapidly swept away leading to a non-axisymmetric inner edge of the circumbinary disc.} 
      \refe{To facilitate the comparison between the}
      typical symmetric case, $q=1$,  and a more strongly asymmetric \refe{mass distribution} case with $q=0.1$, \refe{we show the evolution of the inner edge of the disc 
      with respect to its mean value.} 
      We can see on Fig.~\ref{fig:rcavity} that, while
      the \refe{azimuthally averaged} inner edge of the disc oscillates, \refee{following both the movement of the lump}
      \refe{(which is less noticeable for the $q=0.1$ case)} 
      and the beat frequency with the binary period,
       this overall variation is at most \refee{$\pm 15\%$} 
       and even smaller \refe{along one} 
       binary orbit. 
       \refe{Nevertheless, we also see that the variation of the {azimuthally averaged} inner edge shows wider variation for the $q=0.1$ case. For that reason
       we will use this case as an example in the rest of the paper as it gives us  a wider range of modifications for the  {azimuthally averaged} inner edge, which in turn will lead to 
       a wider range of changes in the SED that we are trying to reproduce.}

\subsection{\refe{Comparing SED from a BBH circumbinary disc with a truncated inner disc around a SBH of the same total 
mass}} 
\label{sec:sed_trunc}

   
       \refe{In  Fig.~10 of \cite{MVC23a}  we showed that} the {\em averaged} energy spectrum of a circumbinary disc, with all its structure, computed in a full BBH metric \refe{could be reproduced}
        by truncating the inner  region of \refe{an axisymmetrical} disc around a {\bf single} \refe{BH} of the total mass of the binary. 
        \refe{Here we want to explore if the} change 
        \refee{at most $\pm 15\%$}  \refe{seen in} the azimuthally averaged inner edge position \refe{translates into a noticeable}
         impact \refe{on} the \refe{instantaneous} energy spectrum of the \refe{entire} circumbinary disc.
  \\ 
     
    \noindent   It is therefore interesting to see if 
        a simple disc around a SBH can reproduce the emission  
       \refe{expected from the circumbinary disc and its associated \lq lump\rq.} 
       For the rest of the paper we will compute observables for a fiducial {BBH} of
        a total mass of $\mathrm{M}_\mathrm{BBH}=10^{6}\, \mathrm{M_\odot}$ \refe{(which is a mass enabling the system to stay in the pre-merger phase for a long enough time for a potential detection, see Fig.~\ref{fig:tmerger})} with a binary separation of $20\, \mathrm{r_g}$, seen at a distance $D=500$~Mpc and under an inclination of $70^\circ$.
\begin{figure}
\includegraphics[width=8.5cm]{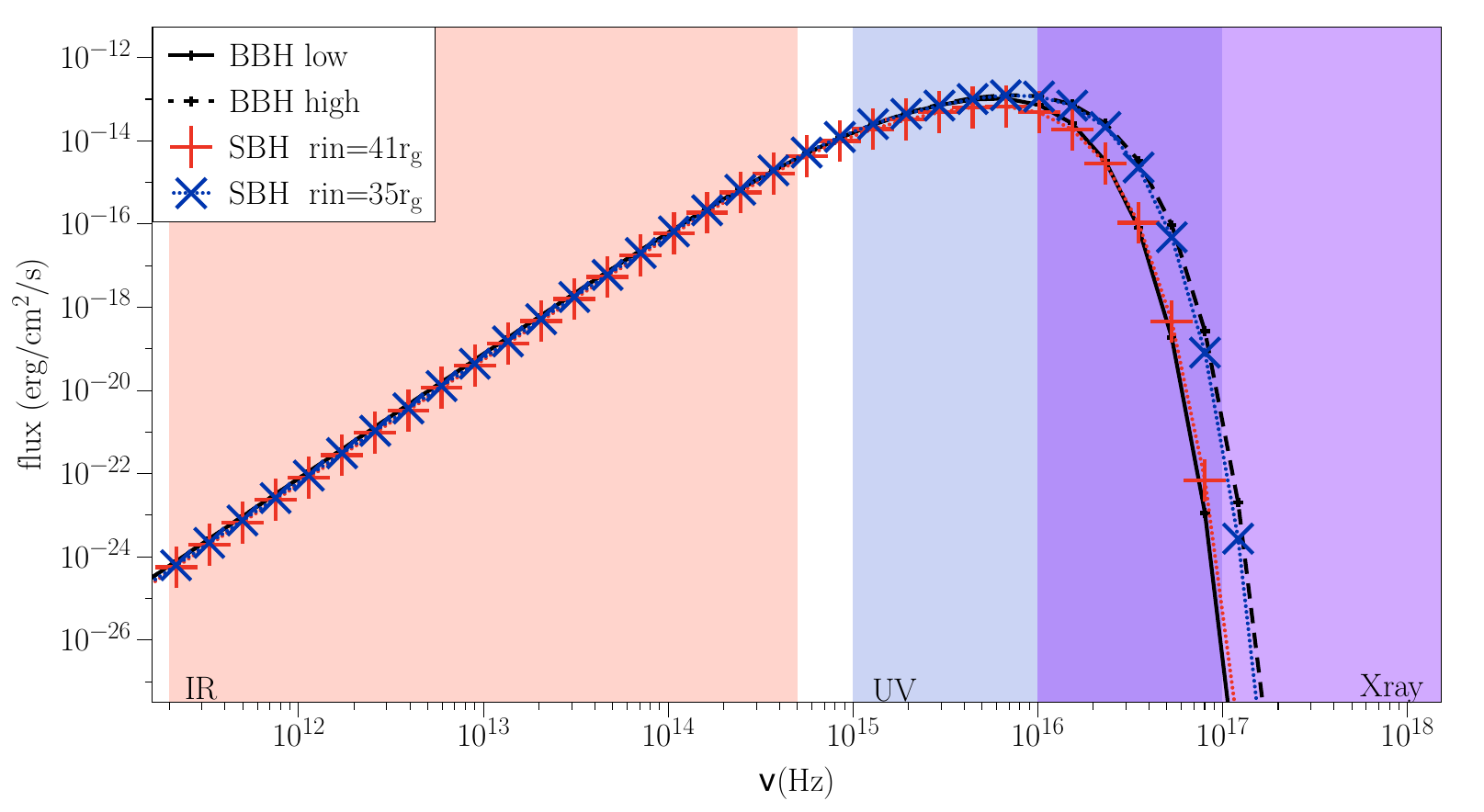}
\caption{Evolution of the energy spectrum along one orbit of the \lq lump\rq\ (the black line for the low point, and dashed  black line for the high point) and how it can be reproduced by a 
moving truncated disc around a single Schwartzchild black hole (blue cross representing an inner edge of the disc at $35\, \mathrm{r_g}$ and red plus $41\, \mathrm{r_g}$). The BBH source is
$\mathrm{M_{BBH}}=10^{6}\, \mathrm{M_\odot}$ with a mass ratio of $q=0.1$ and a separation of $20\, \mathrm{r_g}$ , seen at a distance $D=500$~Mpc and under an inclination of $70^\circ$.
}
\label{fig:spectrum_trunc}
\end{figure}
    We see in Fig.~\ref{fig:spectrum_trunc} that even 
    \refe{for the \refee{$15$\%} variation of the  $q=0.1$ case}, the full energy spectrum can be reproduced by having an
    axisymmetric disc around a  \refe{Schwarzschild} BH \refe{with the same mass} if the inner edge of that disc is further away than its last stable orbit and matches the \refe{average} position of the circumbinary cavity as seen in Fig.~\ref{fig:rcavity}. 
     This is related to the fact that the differences \refe{due to the}  GR effects between a BBH and a SBH of the same total mass are relatively small at that distance \citep[this can be seen in Fig.~3. of ][]{MVC23a}, hence the overall \refe{instantaneous}
     spectra for a disc at the same distance are similar and {\bf do not  allow us to distinguish between the different central objects}.
  \\
     

   
   \noindent \refee{The drawback of that fit is that it leaves the need to explain the reason for such truncation of the disc. On top of that,  if we  have access to more than one spectrum of the system, 
    we will find that the inner edge of the disc is moving which also needs to be explained.}
   
\subsection{\refe{Comparing SED from a BBH circumbinary disc with a disc reaching the last stable orbit of a heavier SBH}} 
\label{sec:heavySED}

    \refe{As a consequence it is interesting to look for the possibility to match the SED of a circumbinary disc with the one produced by a SBH disc extending up to \refee{its}  last stable orbit.}
\\ 
   Following the similitude between the circumbinary spectra and truncated discs shown previously \refe{\cite[also raised by][]{roedig_observational_2014}}, we use the temperature scaling from \cite{shakura73} to estimate its distribution in the circumbinary disc as well.
 Hence, the temperature profile for a black hole of mass $M_\mathrm{BBH}$  with an inner edge of  the disc at $r_\mathrm{in}^\mathrm{BBH}$ 
 is given by
 \refee{
 \be
   T^\mathrm{BBH} (r) \propto   \left( \frac{ \dot{\mathrm{M}}_\mathrm{BBH}}{\dot{\mathrm{M}}_\mathrm{Edd}}\right)^{1/4}  \left(\mathrm{M}_\mathrm{BBH} \right)^{-1/4}    \left(\frac{r}{r_\mathrm{in}^\mathrm{BBH}} \frac{r_\mathrm{in}^\mathrm{BBH}}{r_g^\mathrm{BBH}} \right)^{-3/4} \nonumber
    \ee
 with $r_g^{\mathrm{BBH}}$ the gravitational radius of the binary, $ \dot{\mathrm{M}}_\mathrm{BBH}$ its accretion rate and $\dot{\mathrm{M}}_\mathrm{Edd}$ the Eddington accretion rate.}
   
 \noindent  In order to be able to fit the same SED we need to have the same effective temperature profile but for a disc reaching its last stable orbit.
  This  translates into a relation between the {BBH} total mass, its \refe{ circumbinary disc inner radius and the required mass for a SBH whose disc reaches its last stable orbit, denoted $r_\mathrm{LSO}^\mathrm{SBH}$}, namely\refee{
\be   
\mathrm{M}_\mathrm{SBH}  =  \left( \frac{r_\mathrm{in}^\mathrm{BBH}}{r_g^\mathrm{BBH}}\right)^{3/2}\left(\frac{r_g^\mathrm{SBH}}{r_\mathrm{LSO}^{\mathrm{SBH}}}\right)^{3/2}  \mathrm{M}_\mathrm{BBH}.
\label{eq:temp}
\ee 
  }
  \noindent    \refe{Since we focus on the pre-decoupling phase, $r_\mathrm{in}^\mathrm{BBH}\, {\approx}\, 2 \, b \, {>} \, 20\,  r_\mathrm{g}^\mathrm{BBH}$ for all cases depicted in Fig.~\ref{fig:tmerger}, with $b$ the orbital separation. Similarly, for the SBH case we have\refee{, depending on its spin,} $1   {\leq} \,   r_\mathrm{LSO}^\mathrm{SBH}\,  {\leq}\,  9 \, r_\mathrm{g}^\mathrm{SBH}$
  \refee{, with $r_g^{\rm SBH}$ the gravitational radius of the SBH.}
  \refee{This} 
  gives us  $\mathrm{M}_{\rm SBH}\, {>} \mathrm{M}_{\rm BBH}$.}
  It is interesting to note that the ratio is at the power of three-half which means that even a small, \refee{$\pm 15\%$, change} 
  in the inner radius of the circumbinary disc will require a larger\refee{, but still mostly within typical error bars,} change for the single \refe{BH} mass.
       \noindent    \refe{This is valid for any spin of the SBH but adding  another
      unknown, will only increase the complexity of the fit and would only make sense if we had an independent constraint on the spin of the object we are trying to fit. 
       For that reason, \refe{in the following} we only show results for a Schwarschild SBH but the calculation can be done for any value of $r_{\rm LSO}^{\rm SBH}$.} \\
      
\begin{figure}
\includegraphics[width=8.5cm]{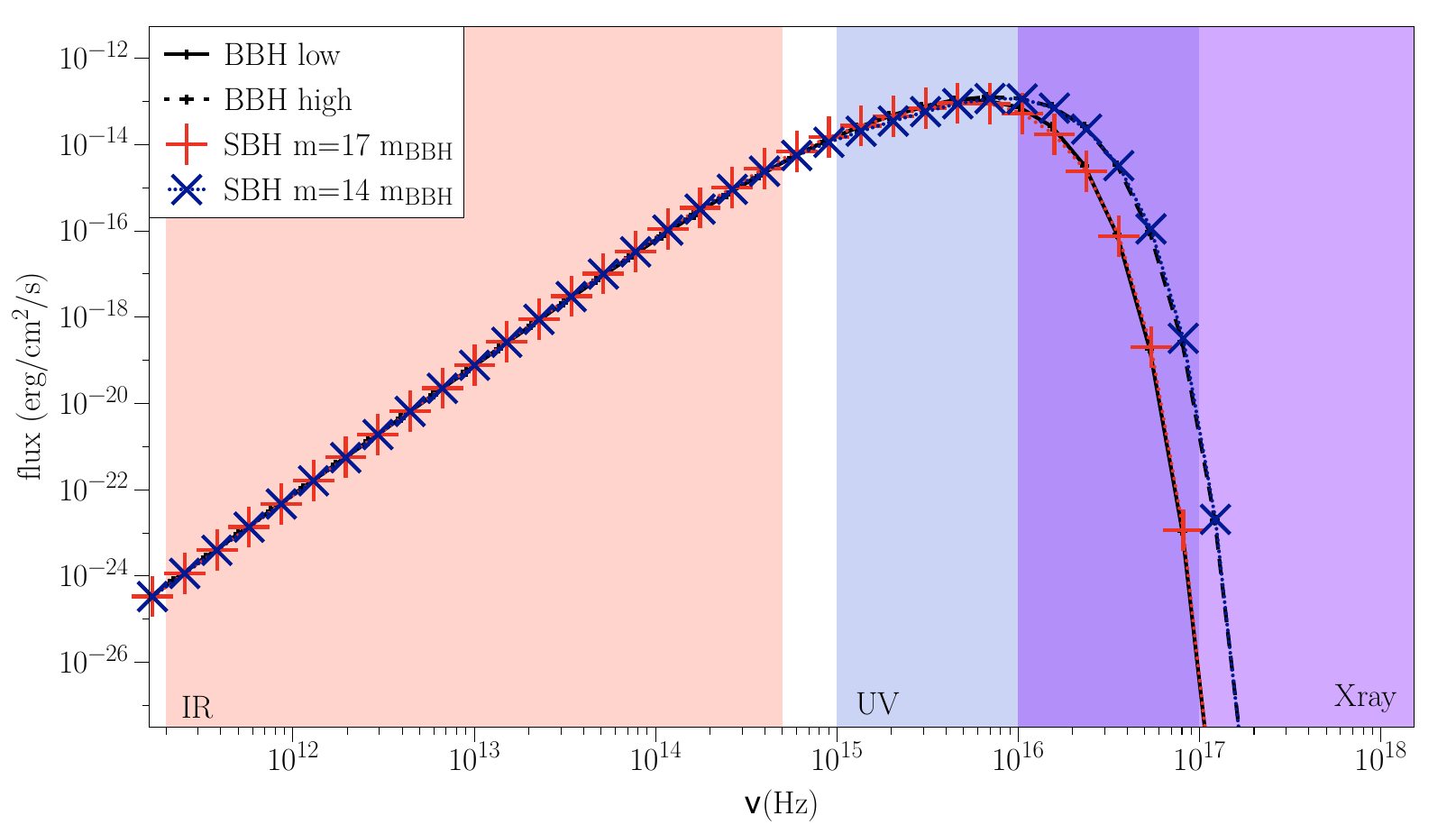}
\caption{Evolution of the energy spectrum along one orbit of the \lq lump\rq\ and how it can be reproduced by a heavier single Schwarzchild \refe{BH} with a disc at its last stable orbit.
The BBH source is $M_\mathrm{BBH}=10^{6}\, \mathrm{M_\odot}$ with a mass ratio of $q=0.1$ and a separation of $20\, \mathrm{r_g}$ , seen at a distance $D=500\,\mathrm{Mpc}$ and under an inclination of $70^\circ$. \refee{An equivalent SBH system whose mass is between $14$ and $17$ times heavier would provide a good fit for the circumbinary disc.}}
\label{fig:spectrum_heavy}
\end{figure}

      When applying this for our previous \refe{circumbinary disc where the inner radius fluctuates} between $34\, r_\mathrm{g}$ and $44 \, r_\mathrm{g}$ we see that it can cause the needed SBH mass to change by a maximum of \refee{$30\%$} along the orbit.
   Nevertheless, when looking at the instantaneous case, it predicts a good fit by a single \refe{BH} \refee{with a mass varying between 14 and 17 times} that of the binary, as is shown in Fig.~\ref{fig:spectrum_heavy}.
   \refee{In the case where we have only one} 
    \refe{instantaneous} energy spectrum of the circumbinary disc, and therefore only one \refe{value for } \refe{its} inner edge \refe{radius}, 
    it will be well approximated by a single, but heavier,  Schwarzchild \refe{BH} with a disc reaching its last stable orbit. 
    \refee{Without any other information, such as an estimate of the central object mass, there is no need for a more complex system than a SBH.}
    \refee{Alternatively,} 
     if we get \refe{a large enough set of} observations to follow the shrinking of the binary orbit, hence \refee{observations} on a long \refee{enough} timescale to 
     \refee{follow the} dynamical \refee{change} of the separation, we could see the needed mass change
    at a rate compatible with Eq.~\ref{eq:temp} \refee{which could be used to hint at the presence of a binary.}
\\

\section{\refe{Reproducing} the temporal variation in BBH: QPOs in SMBH} 
\label{sec:QPO} 
\subsection{Interest of timing studies for BBH candidates}

     \refe{While the SED is the first and most common observable for supermassive BHs, another common observable of BH systems is their time variability, especially when a close to periodic signal
     exists.}
     \refe{Because of the difference in timescale, such studies tend to be more frequent for low-mass X-ray binaries, with a}  
     modulation in the range of seconds to 100 seconds, 
     \refe{while it is } at least hours for the lowest \refe{supermassive BH} 
      and up to hundreds of days
     in the highest mass systems \citep[see for example][]{V23}. Studying such variability is technically challenging and requires long spanning observations to constrain, which is the 
     reason that we have only limited  variability studies in supermassive BHs.\\

     \noindent Nevertheless, looking at the \lq fast\rq\   temporal variability of SMBH is becoming particularly interesting with the rise of the lump \refe{feature} and its variability as a 
     potential signature of near-merger {BBH}s \citep{shi_three-dimensional_2012,noble_circumbinary_2012,dorazio_accretion_2013,farris_binary_2014,noble_mass-ratio_2021-1} 
    among other periodic-like signatures \refeee{\citep{dorazio_observational_2023,franchini_emission_2024}}.     
    \refeee{As we have excised the inner region, close to the black-holes, focusing on the circumbinary disc only, we lack any variability linked to the presence of individual accretion
    structures. While studies of those \lq mini-disc\rq\ exist \citep{dorazio_relativistic_2015,krauth_self-lensing_2024} their variability cannot be easily estimated from  the literature. Nevertheless,
    their flux should be modulated either at the (semi-)orbital period or beat with the lump, which is different from the lump period \citep{WesternacherSchneider22}.  
    Hence, having this region does not change the existence of the lump modulation, though it would change the amplitude ratio between the two modulations 
    (linked with the lump and the orbital period of the binary) depending on the energy band where the lightcurve is computed \citep{WesternacherSchneider22}.  While this last point could be 
    a novel way to detect binary black-holes, we are focusing here on the longer-lasting circumbinary disc \citep{krauth_disappearing_2023}.
    }
     
     At the same time, in order to check the viability of the lump variability as a pre-merger signature we need to search 1) if there is a way to mimic the periodic signal predicted 
     to come from the orbiting lump in a system with only one \refe{BH}, 2) how probable those would be and 3) if there is any way to distinguish between them.  

 \subsection{ Comparing the lightcurve modulations from the lump in BBH systems and from \refee{QPOs} in SBH systems}
 
 
 


Similarly to what was done for the thermal spectrum observable, here we aim to match the lightcurve modulation frequency, induced by the presence of 
the lump in circumbinary discs, to the variability frequency sometimes observed in SBH systems. Indeed SBH accretion discs are known to exhibit \refee{QPOs,} quasi periodic modulation 
of their lightcurve, likely induced by a fluid instability \cite[see e.g.][]{remillard06}.   \\

\begin{figure}
\includegraphics[width=8.5cm]{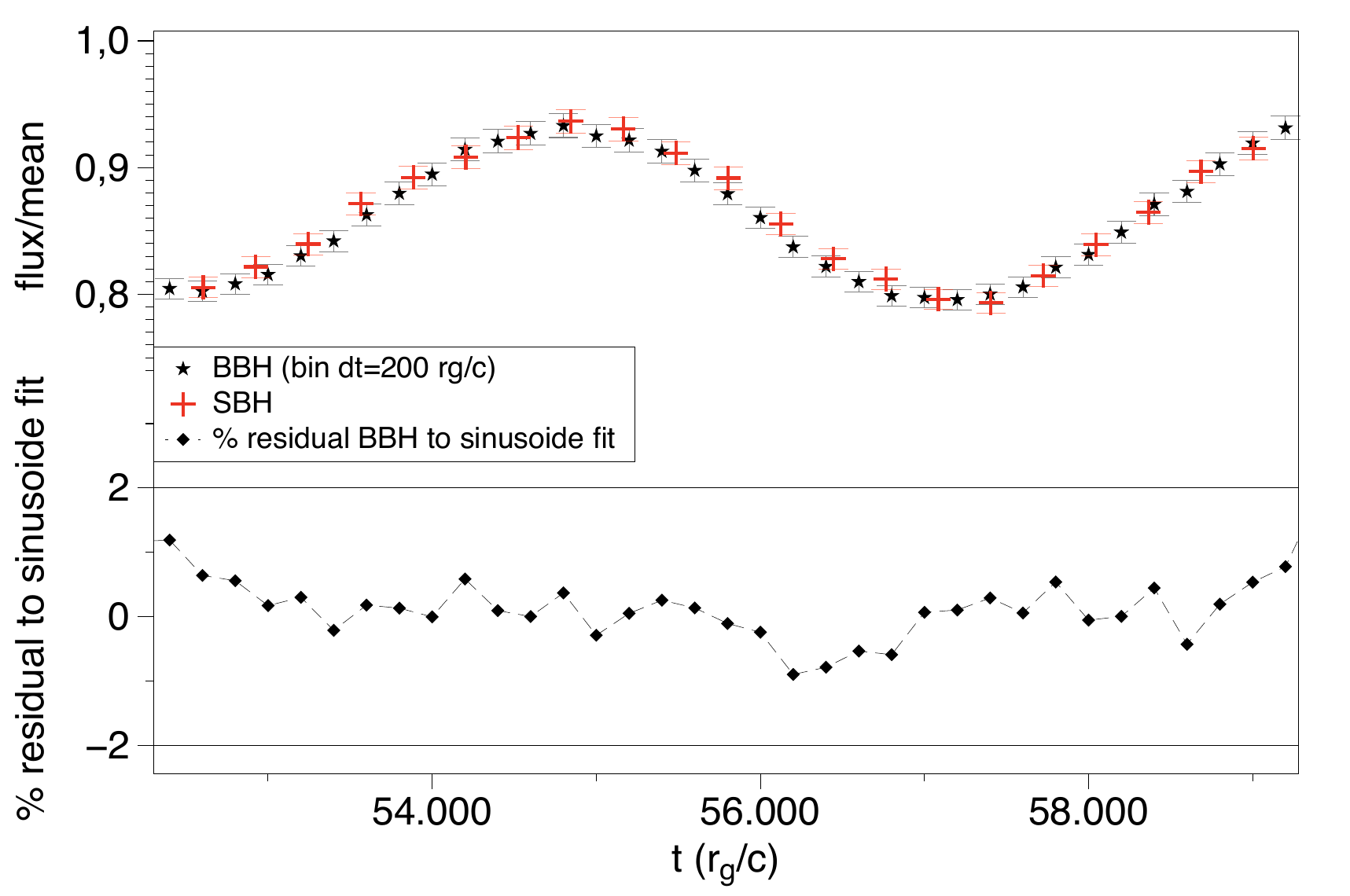}
\caption{Comparing the lightcurve from a SBH \refe{prone to an instability occurring 
at the last stable orbit of its disc} versus a lower mass BBH system exhibiting the so-called \lq lump\rq\ 
at the inner edge of its circumbinary disc. The error bar
corresponds to a one percent error. The bottom panel shows the percent residual of fitting the BBH lightcurve with a simple sinusoid often used to represent the RWI induced periodic variability in black hole 
systems in a model-independent way. Those lightcurves were computed at  an inclination of $70^\circ$.}
\label{fig:RWI_lc}
\end{figure}

      \indent  \refee{Even before having a model to explain the observed variability, we might try if 
a model-independent variability for the SBH case ({\em i.e.} we do not explain its origin) gives an  acceptable fit. 
           Indeed, this might be the first \lq model\rq\  tried on such observations as to not add model-dependent hypotheses \refe{\citep{dorazio_relativistic_2015}}. 
           The bottom of Fig.~\ref{fig:RWI_lc}, shows the residual from fitting the BBH's lightcurve variability with a simple sinusoid \refee{(the full fit can be seen on the top of Fig.~\ref{fig:binnedLC})}.
           Not only does the residuals stay  within a few percent of the data but they also lack any strong additional feature pointing out to a missing component.}
  \\

     \refee{The next option to fit such variable lightcurve would be to have a QPO-inducing instability in a} 
    SBH system with a similar \refee{BBH} mass \refee{and} the inner edge of its accretion disc far from its last stable orbit (so their respective SEDs are similar, 
     Sec.~\ref{sec:sed_trunc}). \refee{This} would be 
     indistinguishable from the circumbinary case as the gravitational field of the central object is unlikely to affect \refee{quantitatively the modulation.} 
         \refee{Nevertheless,} this would, \refee{as in the SED case,} 
     raise the question of what causes this sharply truncated inner disc.
\\

      \refee{Similarly to what was done for the spectrum, another possibility} would be to try to reproduce the observed lightcurve with a SBH having its disc reaching its last stable orbit, 
      hence removing the need for \refe{an} additional mechanism to 
      truncate the inner disc.
       For that we need to see if not only the frequency of the lump \refe{variability} can be matched, but more importantly, how well \refe{the variability induced by an instability, such as the RWI  \citep{V20},}  
       occurring at the edge of a disc for a single \refe{BH} \refe{could be confused with} the 
     lightcurve modulation coming from the lump  at the edge of the circumbinary disc.          
         By equating  the orbital \refe{Keplerian} frequency at the edge of the circumbinary disc with the one at the last stable orbit of a single \refe{Schwarzschild} \refe{BH} we get the requirement for the SBH  to be more massive as  
$\mathrm{M}_\mathrm{SBH} =  \left({r_\mathrm{in}^\mathrm{BBH}}/{r_\mathrm{LSO}^\mathrm{SBH}}\right)^{3/2}   \mathrm{M}_\mathrm{BBH} $  (with both radii given in units of their respective $r_\mathrm{g}$ in order to remove the mass scaling). \refee{Here we see that the needed mass of the SBH inferred from the modulation is the same as the one needed for the SED fitting, meaning that the full SED will
also be reproduced.}

    \refe{Looking at the lightcurves of both systems,} we need to see how well the
    \refe{lightcurve of} such SBH would be able to match the lightcurve from the circumbinary disc. 
    Using {\tt e-NOVAs} and the previous mass estimate we can compute the SBH lightcurve \refe{\citep[with the appropriate mass using simulations from][]{V20}}
    and compare it with the \refe{corresponding} circumbinary lightcurve. \ 
    The top of Fig.~\ref{fig:RWI_lc} shows how little the lightcurve 
     depends on the nature of the central object, indeed, 
    the lightcurve computed from the more massive SBH is easily within the 1\% error bar of the lightcurve computed from the circumbinary simulation.

      \noindent \refee{The similitude of both lightcurves makes it} 
      very difficult to differentiate between a BBH and a heavier SBH purely based on \refee{the fitting of the modulation.} 
      \refee{This will also stay difficult  if we take into account the change in frequency that will occurs has the binary separation decreases.}
      \refe{Indeed, as can be seen on Fig.~\ref{fig:tmerger}, the less massive of the supermassive 
      {B}BHs will go through the inspiral too fast for us to detect the}
      change in 
      frequency  induced by the edge of the circumbinary disc moving inward.
      \refe{Similarly, the higher mass systems will have too slow of an evolution to produce a detectable change in the inner edge of the circumbinary disc.}
      \refe{Only 
      {B}BHs around $10^{6-7} \mathrm{M_\odot}$, if caught early,} 
      might 
      \refee{give us enough observations of the period evolution that the inferred SBH mass for each observation would become incompatible with each other.}

  \section{\refe{A general view on} all the observables together}    

     We have seen that neither the spectrum nor the lightcurve independently can differentiate between a circumbinary or a SBH, here we want to explore  if, and how,  we could push those in order to 
     have enough tension \refe{between the SBH fits to the data to rule that explanation out, hence leaving us with only the BBH explanation.}

   \subsection{Can mass determination help lift the degeneracy? }
   
 
      A way to lift that issue is possible, at least for the larger circumbinary disc\refee{, hence the wider binary}. Indeed the mass needed for the \refee{different fits presented above
      are greater than the  actual mass of the binary by
       $\mathrm{M}_\mathrm{SBH} =  \left({r_\mathrm{in}^\mathrm{BBH}}/{r_\mathrm{LSO}^\mathrm{SBH}}\right)^{3/2}   \mathrm{M}_\mathrm{BBH} $ 
 }
     hence\refee{, for the wider separation,} creating a discrepancy
      that might be too large to be within an acceptable range of mass based on the quality of the SEDs and lightcurves. 
     In the case used to illustrate the fitting \refe{in the previous examples} this leads to a mass difference of  
     \refee{$M^\mathrm{SBH} \simeq 15 M^\mathrm{BBH}$} which could put a strain on the interpretation that \refe{the system hosts} 
      a single \refe{BH}. 
      This might be enough  for the larger circumbinary disc\refe{, i.e. binary with larger \refee{separation},} 
      if we also  have access  to a  good and {\bf independent} mass determination  of the system as fitting the SED with a SBH at its last stable orbit requires a 
     black hole that could be more than one order of magnitude heavier than the BBH is. 
   \\

     Nevertheless, if we lift the requirement for the disc to be at its last stable orbit then it will be impossible to distinguish a circumbinary system from a \{SBH + truncated disc\}
      system as the metric at such distance is too similar and there is little difference in the behaviour of the instability. 
      This would raise the question of what causes the receding of the  inner disc  but would give a great fit of the data and be hard to refute using only the data presented here.

 \subsection{Detecting the binary period}

\begin{figure}
\centering
\includegraphics[width=8cm]{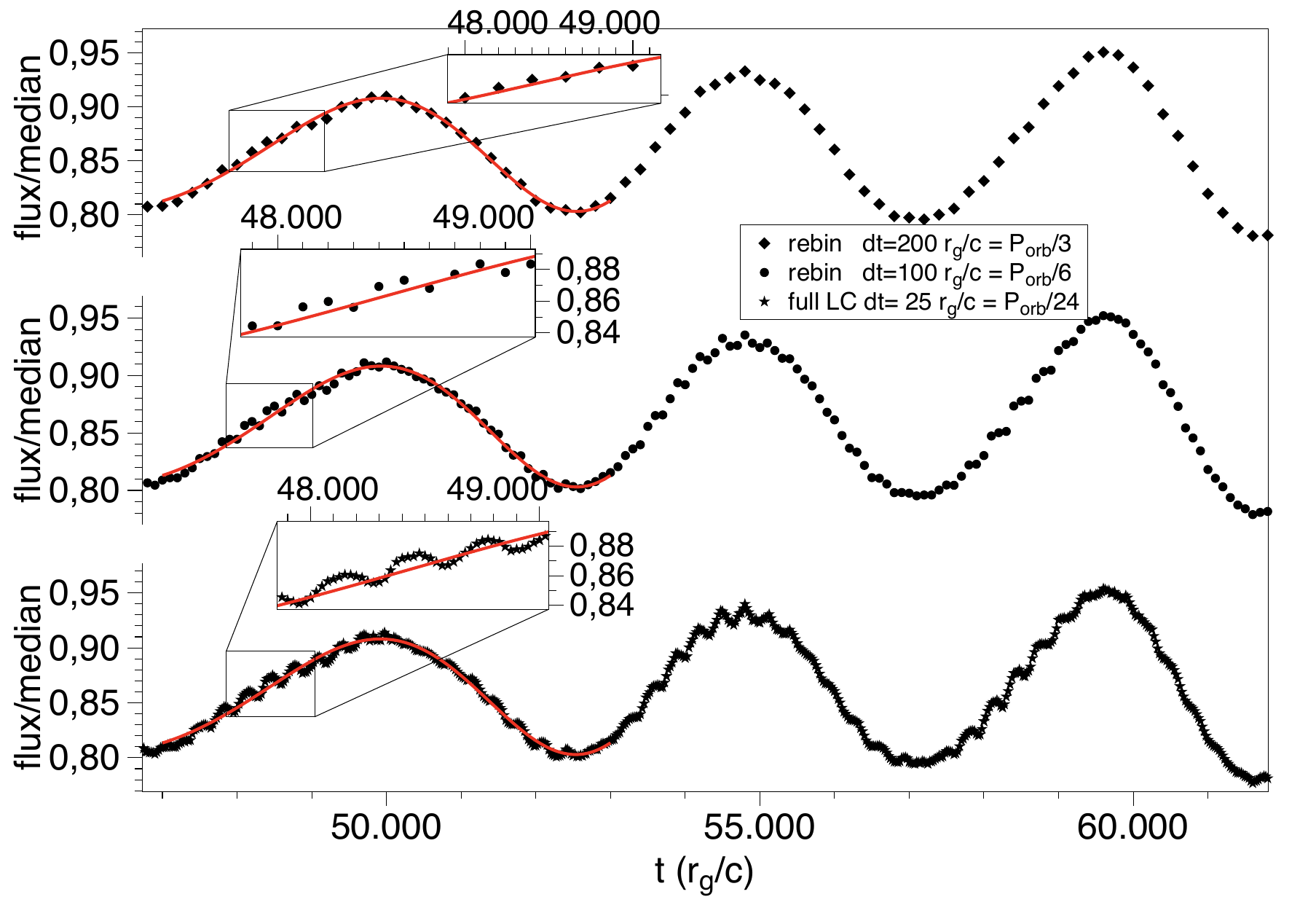}  
\caption{Lightcurve for the BBH simulation binned at different time resolutions, $P_\mathrm{orb}/3$, $P_\mathrm{orb}/6$ and $P_\mathrm{orb}/24$, showing the need for a good time resolution and signal-to-noise ratio 
in order to identify both frequencies. Those lightcurves were computed at an inclination of $70^\circ$.}
\label{fig:binnedLC}
\end{figure}

           We now turn to look how improving the data could help distinguish between a single \refe{BH} and a binary, focusing on what is unique to the binary, namely 
           its period \refe{and features related to it, such as the spiral density waves and the binary-lump beat, both producing a signal close to the (semi-) binary period}. 
 \\          
   Following this idea, we binned the lightcurve for the BBH simulation at different time resolutions, $P_\mathrm{orb}/3$, $P_\mathrm{orb}/6$ and $P_\mathrm{orb}/24$. 
    \refe{As the BBH 
   rotates at a larger frequency than the plasma in the circumbinary disc, the binary modulation is expected to have a smaller period \refee{than} the lump induced modulation of the disc.} 
    Fig.~\ref{fig:binnedLC} shows how we can see half the binary period directly on top of the main variability once we have enough data points per binary period.
    The red curve on each LC represents  a simple sinusoidal fit of the main modulation and the zoom-in plots give an idea of how that secondary modulation would become 
    visible in the residual 
    if a model-independent sinusoide was used to fit the LC.  
       It is important to note that, while a good time resolution is necessary, it is not enough as the \refe{lower-amplitude} modulation is \refe{only} of the order of $1\%$ of the total flux 
       \refe{for a wide range of separations}, hence much smaller than the lump's one and 
   will be hard to detect without a good signal-to-noise ratio for the lightcurve. 
   \refeee{It is worth noting, that this secondary modulation might be strengthen if also present in the individual accretion structure of each black-hole, though we
   will see an evolution of the respective amplitude depending on the energy band where the LC is computed \citep{WesternacherSchneider22}.}
\\

   While being able to directly observe the impact of the binary period on the lightcurve will be a great step toward distinguishing between a single \refe{BH} system and a binary system, it does not 
   create a unique enough impact on the lightcurve. 
   Indeed, such small modulation seen on top of a stronger, lower frequency, modulation is similar to what is seen in low-mass X-ray binaries in the state where both the low-frequency and high-frequency 
   {quasi-periodic oscillations (LF and HF QPOs)} are present. This is especially important as most models for those QPOs could be applied to SMBH \citep[see for example][]{VTR11,V23} and would reach a similar frequency zone in the late stage of merger. 
 \begin{figure}
\centering
\includegraphics[width=8.5cm]{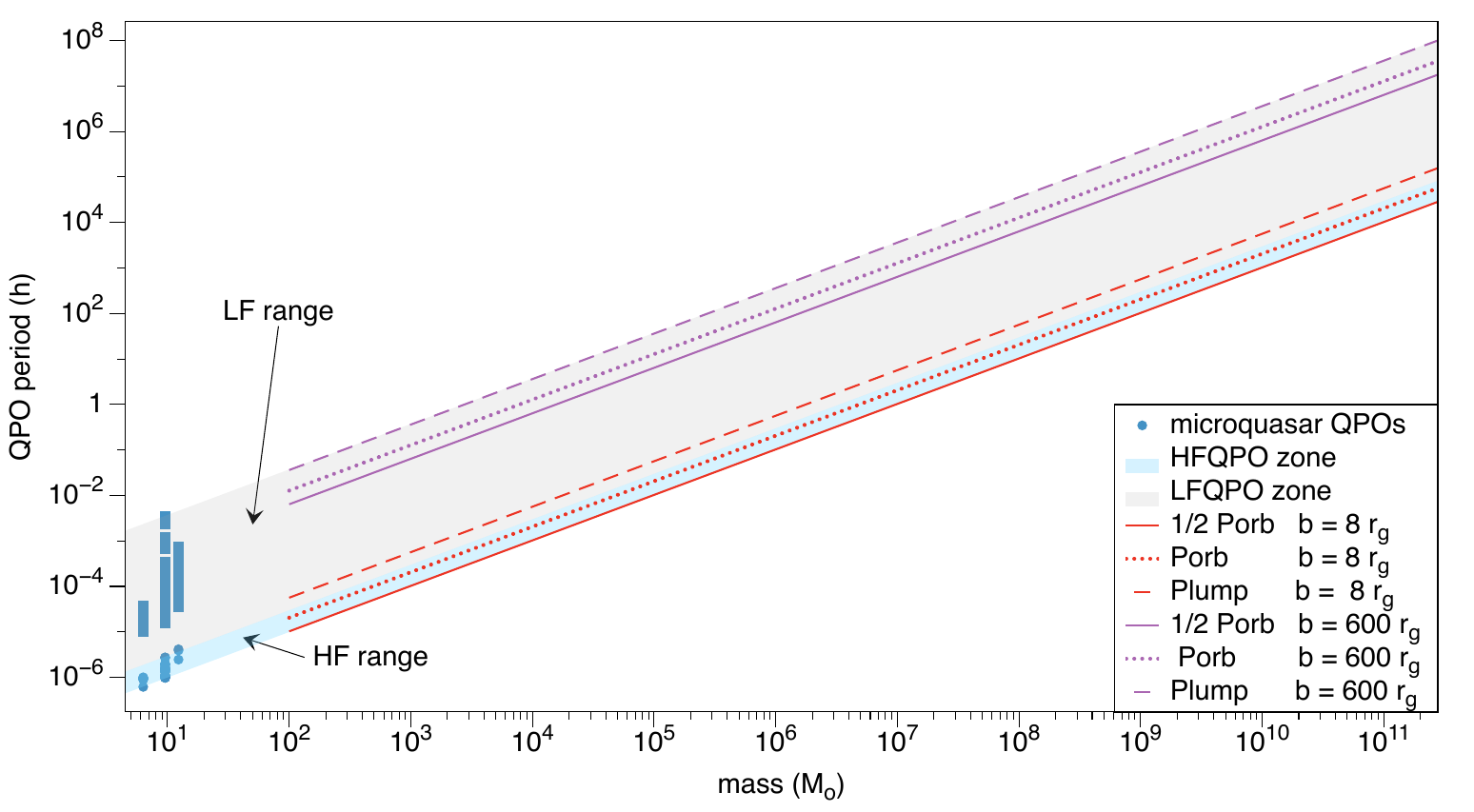}  
\caption{Range of modulation periods as a function of mass for the circumbinary modulation, namely the lump and binary period \refe{(also indicative of the binary-lump beat period)} 
and how they compare to an extension of the LF and HF QPO frequency of 
low mass X-ray binaries \citep[data for GRO J$1655-40$, GRS $1915+105$ and XTE J$1550-564$ from][]{rodriguez02,mikles09,VV17}. The $600\, r_\mathrm{g}$ separation is to show that all binary separations
with a disc will fall into the QPO range.} 
\label{fig:QPOs}
\end{figure}

   Indeed, if one extends the range of observed low-frequency and high-frequency QPOs to higher masses using a simple mass scaling \cite[see for example][]{STW21} we see on Fig.~\ref{fig:QPOs} that the
   expected lump and binary period will be in the \lq QPO range\rq\  for a wide variety of orbital separations\footnote{The upper limit of a binary separation of $600\, r_\mathrm{g}$ is 
   only there to show what would be the equivalent to the upper limit of LFQPOs and not as a realistic limit for the existence of a circumbinary disc.}   {\bf making the presence of both variabilities, 
   associated with the lump and the binary period, a necessary condition for a BBH but not sufficient} 
   as they could be originating from instabilities 
   or phenomena similar to the case in low-mass X-ray binaries.

\section{conclusion}

    \refee{In this paper we explored the difficulty of identifying if a source hosts a binary system or a single black-hole using either/both the SED and lightcurve from their disc's
    thermal emission without apriori on its nature. The aim
    was to show the potential issues and pitfalls of disentangling  between BBH and SBH based only on their disc's emission and to show potential steps to resolved that.}
    \refee{Indeed, while} 
    the periodic modulation predicted to originate from the orbiting lump at the edge of a circumbinary disc is often presented as a way to detect pre-merger {BBH}s,
    it raises the question of our capacity to distinguish between the \refe{lump induced modulation of a circumbinary disc and a SBH disc exhibiting temporal variability due to the presence of a fluid instability (such as the RWI) occurring near its inner edge}.
    In this paper we show that the modulation caused by the lump will be hard to use as the ultimate signature of BBHs. Indeed, it requires a good, independent, mass determination, as well as 
    a precise time resolution and a good signal-to-noise ratio in the wavelength where the lump modulates the flux, while not being unique to BBH systems.
    Nevertheless, it can be used as a first criteria when looking for {BBH}s. Indeed, while not unique to those systems, its presence is a requirement for a large number 
    of mass ratio and pre-merger times \citep{MVC23b} and can help reduce the number of sources that could be {BBH} candidates.\\

 \noindent    It is interesting to note that what limits the usage of this  double variability as a unique signature of BBH is our limited knowledge of supermassive \refe{single} BH QPOs which prevents us 
    from quantifying the differences between both variabilities.  Hence, we need to improve our  understanding of supermassive \refe{single}  BH QPOs at the same time as we explore the variability 
    linked with the presence of a BBH if we want to be able to differentiate between them in the near future. \\ 
  \noindent      Similarly, the spectra could not be used to uniquely point toward a {BBH}, even in case of a good, independent, mass determination, 
     but it is worth noting that the existence of a truncated disc could also be used to trigger more thorough observations to look for the double variability.
      \refe{Though, if the separation is wide enough, some accretion structures could be present closer to each black-hole, hence {\em filling} some of that cavity making 
      it harder to detect its presence.}
      
      Hence, we see that while neither timing nor spectral features of circumbinary discs are uniquely linked with the presence of a binary, 
      they would be useful in weeding out false binary candidates
      and should be used as a test on all existing candidate-binaries.
       


\section*{Acknowledgements}

RMR acknowledges funding from Centre National d'Etudes Spatiales (CNES) through a postdoctoral fellowship (2021-2023). 
RMR has received funding from the European Research Council (ERC) under the European Unions Horizon 2020 research and innovation programme (grant agreement No. 101002352, PI: M. Linares).
This work was supported by CNES, focused on Athena, the LabEx UnivEarthS, ANR-10- LABX-0023 and ANR-18-IDEX-000, and by the \lq Action Incitative: Ondes gravitationnelles et objets compacts{\rq} and the Conseil Scientifique de l'Observatoire de Paris. 
The numerical simulations we have presented in this paper were produced on the platform DANTE (AstroParticule \& Cosmologie, Paris, France) 
and on the high-performance computing resources from Grand Equipement National de Calcul Intensif (GENCI) - Centre Informatique National de l'Enseignement Sup\'erieur (CINES, grant A0100412463)
 and IDRIS (grants A0130412463 and A0150412463). 

\section*{Data Availability}

The data that support the findings of this study are available from the corresponding author, P.V., on request and will also be part of a data release in 2025\footnote{Which will be 
available for download at
\url{https://apc.u-paris.fr/~pvarni/eNOVAs/LCspec.html}}.

\bibliographystyle{mnras}

\bibliography{BiblioFull}

\begin{thebibliography}{}
\makeatletter
\relax
\def\mn@urlcharsother{\let\do\@makeother \do\$\do\&\do\#\do\^\do\_\do\%\do\~}
\def\mn@doi{\begingroup\mn@urlcharsother \@ifnextchar [ {\mn@doi@}
  {\mn@doi@[]}}
\def\mn@doi@[#1]#2{\def\@tempa{#1}\ifx\@tempa\@empty \href
  {http://dx.doi.org/#2} {doi:#2}\else \href {http://dx.doi.org/#2} {#1}\fi
  \endgroup}
\def\mn@eprint#1#2{\mn@eprint@#1:#2::\@nil}
\def\mn@eprint@arXiv#1{\href {http://arxiv.org/abs/#1} {{\tt arXiv:#1}}}
\def\mn@eprint@dblp#1{\href {http://dblp.uni-trier.de/rec/bibtex/#1.xml}
  {dblp:#1}}
\def\mn@eprint@#1:#2:#3:#4\@nil{\def\@tempa {#1}\def\@tempb {#2}\def\@tempc
  {#3}\ifx \@tempc \@empty \let \@tempc \@tempb \let \@tempb \@tempa \fi \ifx
  \@tempb \@empty \def\@tempb {arXiv}\fi \@ifundefined
  {mn@eprint@\@tempb}{\@tempb:\@tempc}{\expandafter \expandafter \csname
  mn@eprint@\@tempb\endcsname \expandafter{\@tempc}}}

\bibitem[\protect\citeauthoryear{Abbott et~al.,}{Abbott
  et~al.}{2017}]{abbott_gravitational_2017}
Abbott B.~P.,  et~al., 2017, \mn@doi [ApJ] {10.3847/2041-8213/aa920c}, 848, L13

\bibitem[\protect\citeauthoryear{Armengol et~al.,}{Armengol
  et~al.}{2021}]{armengol_circumbinary_2021}
Armengol F. G.~L.,  et~al., 2021, arXiv:2102.00243 [astro-ph]

\bibitem[\protect\citeauthoryear{Armitage \& Natarajan}{Armitage \&
  Natarajan}{2002}]{armitage_accretion_2002}
Armitage P.~J.,  Natarajan P.,  2002, \mn@doi [The Astrophysical Journal]
  {10.1086/339770}, 567, L9

\bibitem[\protect\citeauthoryear{Artymowicz \& Lubow}{Artymowicz \&
  Lubow}{1994}]{artymowicz_dynamics_1994}
Artymowicz P.,  Lubow S.~H.,  1994, \mn@doi [ApJ] {10.1086/173679}, 421, 651

\bibitem[\protect\citeauthoryear{{Casse} \& {Varniere}}{{Casse} \&
  {Varniere}}{2018}]{CV18}
{Casse} F.,  {Varniere} P.,  2018, \mn@doi [\mnras] {10.1093/mnras/sty2475},
  \href {http://cdsads.u-strasbg.fr/abs/2018MNRAS.481.2736C} {481, 2736}

\bibitem[\protect\citeauthoryear{{Cocchiararo}, {Franchini}, {Lupi}  \&
  {Sesana}}{{Cocchiararo} et~al.}{2024}]{2024Cocchiararo}
{Cocchiararo} F.,  {Franchini} A.,  {Lupi} A.,   {Sesana} A.,  2024, \mn@doi
  [arXiv e-prints] {10.48550/arXiv.2402.05175}, \href
  {https://ui.adsabs.harvard.edu/abs/2024arXiv240205175C} {p. arXiv:2402.05175}

\bibitem[\protect\citeauthoryear{D'Orazio \& Charisi}{D'Orazio \&
  Charisi}{2023}]{dorazio_observational_2023}
D'Orazio D.~J.,  Charisi M.,  2023, Observational {Signatures} of
  {Supermassive} {Black} {Hole} {Binaries}, \url
  {http://arxiv.org/abs/2310.16896}

\bibitem[\protect\citeauthoryear{D'Orazio, Haiman  \& MacFadyen}{D'Orazio
  et~al.}{2013}]{dorazio_accretion_2013}
D'Orazio D.~J.,  Haiman Z.,   MacFadyen A.,  2013, \mn@doi [Monthly Notices of
  the Royal Astronomical Society] {10.1093/mnras/stt1787}, 436, 2997

\bibitem[\protect\citeauthoryear{D'Orazio, Haiman  \& Schiminovich}{D'Orazio
  et~al.}{2015}]{dorazio_relativistic_2015}
D'Orazio D.~J.,  Haiman Z.,   Schiminovich D.,  2015, \mn@doi [Nature]
  {10.1038/nature15262}, 525, 351

\bibitem[\protect\citeauthoryear{Duffell, D’Orazio, Derdzinski, Haiman,
  MacFadyen, Rosen  \& Zrake}{Duffell et~al.}{2020}]{duffell_circumbinary_2020}
Duffell P.~C.,  D’Orazio D.,  Derdzinski A.,  Haiman Z.,  MacFadyen A.,
  Rosen A.~L.,   Zrake J.,  2020, \mn@doi [ApJ] {10.3847/1538-4357/abab95},
  901, 25

\bibitem[\protect\citeauthoryear{{Duffell} et~al.,}{{Duffell}
  et~al.}{2024}]{codecomp_SB}
{Duffell} P.~C.,  et~al., 2024, \mn@doi [arXiv e-prints]
  {10.48550/arXiv.2402.13039}, \href
  {https://ui.adsabs.harvard.edu/abs/2024arXiv240213039D} {p. arXiv:2402.13039}

\bibitem[\protect\citeauthoryear{D’Ascoli, Noble, Bowen, Campanelli, Krolik
  \& Mewes}{D’Ascoli et~al.}{2018}]{dascoli_electromagnetic_2018}
D’Ascoli S.,  Noble S.~C.,  Bowen D.~B.,  Campanelli M.,  Krolik J.~H.,
  Mewes V.,  2018, \mn@doi [ApJ] {10.3847/1538-4357/aad8b4}, 865, 140

\bibitem[\protect\citeauthoryear{Farris, Duffell, MacFadyen  \& Haiman}{Farris
  et~al.}{2014}]{farris_binary_2014}
Farris B.~D.,  Duffell P.,  MacFadyen A.~I.,   Haiman Z.,  2014, \mn@doi [ApJ]
  {10.1088/0004-637X/783/2/134}, 783, 134

\bibitem[\protect\citeauthoryear{Farris, Duffell, MacFadyen  \& Haiman}{Farris
  et~al.}{2015}]{farris_characteristic_2015}
Farris B.~D.,  Duffell P.,  MacFadyen A.~I.,   Haiman Z.,  2015, \mn@doi
  [MNRAS] {10.1093/mnrasl/slu160}, 446, L36

\bibitem[\protect\citeauthoryear{Franchini, Bonetti, Lupi  \& Sesana}{Franchini
  et~al.}{2024}]{franchini_emission_2024}
Franchini A.,  Bonetti M.,  Lupi A.,   Sesana A.,  2024, Emission signatures
  from sub-pc {Post}-{Newtonian} binaries embedded in circumbinary discs, \url
  {http://arxiv.org/abs/2401.10331}

\bibitem[\protect\citeauthoryear{{Gold}}{{Gold}}{2019}]{gold_relativistic_2019}
{Gold} R.,  2019, Galaxies

\bibitem[\protect\citeauthoryear{Gold, Paschalidis, Etienne, Shapiro  \&
  Pfeiffer}{Gold et~al.}{2014}]{gold_accretion_2014}
Gold R.,  Paschalidis V.,  Etienne Z.~B.,  Shapiro S.~L.,   Pfeiffer H.~P.,
  2014, \mn@doi [Phys. Rev. D] {10.1103/PhysRevD.89.064060}, 89, 064060

\bibitem[\protect\citeauthoryear{Graham et~al.,}{Graham
  et~al.}{2020}]{graham_candidate_2020}
Graham M.~J.,  et~al., 2020, \mn@doi [Phys. Rev. Lett.]
  {10.1103/PhysRevLett.124.251102}, 124, 251102

\bibitem[\protect\citeauthoryear{Gutiérrez, Combi, Noble, Campanelli, Krolik,
  López~Armengol  \& García}{Gutiérrez
  et~al.}{2022}]{gutierrez_electromagnetic_2022}
Gutiérrez E.~M.,  Combi L.,  Noble S.~C.,  Campanelli M.,  Krolik J.~H.,
  López~Armengol F.,   García F.,  2022, \mn@doi [ApJ]
  {10.3847/1538-4357/ac56de}, 928, 137

\bibitem[\protect\citeauthoryear{Ireland, Mundim, Nakano  \&
  Campanelli}{Ireland et~al.}{2016}]{ireland16}
Ireland B.,  Mundim B.~C.,  Nakano H.,   Campanelli M.,  2016, \mn@doi [PRD]
  {10.1103/PhysRevD.93.104057}, 93

\bibitem[\protect\citeauthoryear{{Krauth}, {Davelaar}, {Haiman},
  {Westernacher-Schneider}, {Zrake}  \& {MacFadyen}}{{Krauth}
  et~al.}{2023}]{krauth_disappearing_2023}
{Krauth} L.~M.,  {Davelaar} J.,  {Haiman} Z.,  {Westernacher-Schneider} J.~R.,
  {Zrake} J.,   {MacFadyen} A.,  2023, \mn@doi [\mnras]
  {10.1093/mnras/stad3095}, \href
  {https://ui.adsabs.harvard.edu/abs/2023MNRAS.526.5441K} {526, 5441}

\bibitem[\protect\citeauthoryear{Krauth, Davelaar, Haiman,
  Westernacher-Schneider, Zrake  \& MacFadyen}{Krauth
  et~al.}{2024}]{krauth_self-lensing_2024}
Krauth L.~M.,  Davelaar J.,  Haiman Z.,  Westernacher-Schneider J.~R.,  Zrake
  J.,   MacFadyen A.,  2024, \mn@doi [Phys. Rev. D]
  {10.1103/PhysRevD.109.103014}, 109, 103014

\bibitem[\protect\citeauthoryear{{Lai} \& {Mu{\~n}oz}}{{Lai} \&
  {Mu{\~n}oz}}{2023}]{LaiMunoz23}
{Lai} D.,  {Mu{\~n}oz} D.~J.,  2023, \mn@doi [\araa]
  {10.1146/annurev-astro-052622-022933}, \href
  {https://ui.adsabs.harvard.edu/abs/2023ARA&A..61..517L} {61, 517}

\bibitem[\protect\citeauthoryear{{Li}, {Finn}, {Lovelace}  \& {Colgate}}{{Li}
  et~al.}{2000}]{Li00}
{Li} H.,  {Finn} J.~M.,  {Lovelace} R.~V.~E.,   {Colgate} S.~A.,  2000, \mn@doi
  [\apj] {10.1086/308693}, \href
  {http://cdsads.u-strasbg.fr/abs/2000ApJ...533.1023L} {533, 1023}

\bibitem[\protect\citeauthoryear{Liu}{Liu}{2021}]{liu_evolution_2021}
Liu W.,  2021, \mn@doi [Monthly Notices of the Royal Astronomical Society]
  {10.1093/mnras/stab1022}, 504, 1473

\bibitem[\protect\citeauthoryear{{Lovelace} \& {Hohlfeld}}{{Lovelace} \&
  {Hohlfeld}}{1978}]{Lovelace78}
{Lovelace} R.~V.~E.,  {Hohlfeld} R.~G.,  1978, \mn@doi [\apj] {10.1086/156004},
  \href {http://cdsads.u-strasbg.fr/abs/1978ApJ...221...51L} {221, 51}

\bibitem[\protect\citeauthoryear{{Lovelace}, {Li}, {Colgate}  \&
  {Nelson}}{{Lovelace} et~al.}{1999}]{Lovelace99}
{Lovelace} R.~V.~E.,  {Li} H.,  {Colgate} S.~A.,   {Nelson} A.~F.,  1999,
  \mn@doi [\apj] {10.1086/306900}, \href
  {http://cdsads.u-strasbg.fr/abs/1999ApJ...513..805L} {513, 805}

\bibitem[\protect\citeauthoryear{MacFadyen \& Milosavljevi{\'c}}{MacFadyen \&
  Milosavljevi{\'c}}{2008}]{macfadyen_eccentric_2008}
MacFadyen A.~I.,  Milosavljevi{\'c} M.,  2008, \mn@doi [ApJ] {10.1086/523869},
  672, 83

\bibitem[\protect\citeauthoryear{Mangiagli, Caprini, Volonteri, Marsat,
  Vergani, Tamanini  \& Inchausp\'e}{Mangiagli
  et~al.}{2022}]{mangiagli_massive_2022}
Mangiagli A.,  Caprini C.,  Volonteri M.,  Marsat S.,  Vergani S.,  Tamanini
  N.,   Inchausp\'e H.,  2022, \mn@doi [Phys. Rev. D]
  {10.1103/PhysRevD.106.103017}, 106, 103017

\bibitem[\protect\citeauthoryear{{Mignon-Risse}, {Varniere}  \&
  {Casse}}{{Mignon-Risse} et~al.}{2023a}]{MVC23a}
{Mignon-Risse} R.,  {Varniere} P.,   {Casse} F.,  2023a, \mn@doi [\mnras]
  {10.1093/mnras/stac379410.48550/arXiv.2212.12005}, \href
  {https://ui.adsabs.harvard.edu/abs/2023MNRAS.519.2848M} {519, 2848}

\bibitem[\protect\citeauthoryear{{Mignon-Risse}, {Varniere}  \&
  {Casse}}{{Mignon-Risse} et~al.}{2023b}]{MVC23b}
{Mignon-Risse} R.,  {Varniere} P.,   {Casse} F.,  2023b, \mn@doi [\mnras]
  {10.1093/mnras/stad177}, \href
  {https://ui.adsabs.harvard.edu/abs/2023MNRAS.tmp..223M} {520, 1285}


\bibitem[\protect\citeauthoryear{{Mikles}, {Varniere}, {Eikenberry},
  {Rodriguez}  \& {Rothstein}}{{Mikles} et~al.}{2009}]{mikles09}
{Mikles} V.~J.,  {Varniere} P.,  {Eikenberry} S.~S.,  {Rodriguez} J.,
  {Rothstein} D.,  2009, \mn@doi [\apjl] {10.1088/0004-637X/694/2/L132}, \href
  {http://cdsads.u-strasbg.fr/abs/2009ApJ...694L.132M} {694, L132}

\bibitem[\protect\citeauthoryear{Noble, Mundim, Nakano, Krolik, Campanelli,
  Zlochower  \& Yunes}{Noble et~al.}{2012}]{noble_circumbinary_2012}
Noble S.~C.,  Mundim B.~C.,  Nakano H.,  Krolik J.~H.,  Campanelli M.,
  Zlochower Y.,   Yunes N.,  2012, \mn@doi [The Astrophysical Journal]
  {10.1088/0004-637X/755/1/51}, 755, 51

\bibitem[\protect\citeauthoryear{Noble, Krolik, Campanelli, Zlochower, Mundim,
  Nakano  \& Zilh{\~a}o}{Noble et~al.}{2021}]{noble_mass-ratio_2021-1}
Noble S.~C.,  Krolik J.~H.,  Campanelli M.,  Zlochower Y.,  Mundim B.~C.,
  Nakano H.,   Zilh{\~a}o M.,  2021, \mn@doi [ApJ] {10.3847/1538-4357/ac2229},
  922, 175

\bibitem[\protect\citeauthoryear{Peters}{Peters}{1964}]{peters_gravitational_1964}
Peters P.~C.,  1964, \mn@doi [Phys. Rev.] {10.1103/PhysRev.136.B1224}, 136,
  B1224

\bibitem[\protect\citeauthoryear{{Remillard} \& {McClintock}}{{Remillard} \&
  {McClintock}}{2006}]{remillard06}
{Remillard} R.~A.,  {McClintock} J.~E.,  2006, \mn@doi [\araa]
  {10.1146/annurev.astro.44.051905.092532}, \href
  {http://adsabs.harvard.edu/abs/2006ARA\%26A..44...49R} {44, 49}

\bibitem[\protect\citeauthoryear{{Rodriguez}, {Varni{e}re}, {Tagger}  \&
  {Durouchoux}}{{Rodriguez} et~al.}{2002}]{rodriguez02}
{Rodriguez} J.,  {Varni{e}re} P.,  {Tagger} M.,   {Durouchoux} P.,  2002,
  \mn@doi [\aap] {10.1051/0004-6361:20000524}, \href
  {http://cdsads.u-strasbg.fr/abs/2002A%26A...387..487R} {387, 487}

\bibitem[\protect\citeauthoryear{Roedig, Krolik  \& Miller}{Roedig
  et~al.}{2014}]{roedig_observational_2014}
Roedig C.,  Krolik J.~H.,   Miller M.~C.,  2014, \mn@doi [ApJ]
  {10.1088/0004-637X/785/2/115}, 785, 115

\bibitem[\protect\citeauthoryear{{Shakura} \& {Sunyaev}}{{Shakura} \&
  {Sunyaev}}{1973}]{shakura73}
{Shakura} N.~I.,  {Sunyaev} R.~A.,  1973, \aap, \href
  {http://adsabs.harvard.edu/cgi-bin/nph-bib_query?bibcode=1973A%26A....24..337S&db_key=AST}
  {24, 337}

\bibitem[\protect\citeauthoryear{Shi \& Krolik}{Shi \&
  Krolik}{2015}]{shi_three-dimensional_2015}
Shi J.-M.,  Krolik J.~H.,  2015, \mn@doi [ApJ] {10.1088/0004-637X/807/2/131},
  807, 131

\bibitem[\protect\citeauthoryear{Shi, Krolik, Lubow  \& Hawley}{Shi
  et~al.}{2012}]{shi_three-dimensional_2012}
Shi J.-M.,  Krolik J.~H.,  Lubow S.~H.,   Hawley J.~F.,  2012, \mn@doi [ApJ]
  {10.1088/0004-637X/749/2/118}, 749, 118

\bibitem[\protect\citeauthoryear{{Smith}, {Tandon}  \& {Wagoner}}{{Smith}
  et~al.}{2021}]{STW21}
{Smith} K.~L.,  {Tandon} C.~R.,   {Wagoner} R.~V.,  2021, \mn@doi [\apj]
  {10.3847/1538-4357/abc9b7}, \href
  {https://ui.adsabs.harvard.edu/abs/2021ApJ...906...92S} {906, 92}

\bibitem[\protect\citeauthoryear{Tang, Haiman  \& MacFadyen}{Tang
  et~al.}{2018}]{tang_late_2018}
Tang Y.,  Haiman Z.,   MacFadyen A.,  2018, \mn@doi [MNRAS]
  {10.1093/mnras/sty423}, 476, 2249

\bibitem[\protect\citeauthoryear{Tiede, Zrake, MacFadyen  \& Haiman}{Tiede
  et~al.}{2021}]{tiede_how_2021}
Tiede C.,  Zrake J.,  MacFadyen A.,   Haiman Z.,  2021, arXiv:2111.04721
  [astro-ph]

\bibitem[\protect\citeauthoryear{Varniere}{Varniere}{2023}]{V23}
Varniere P.,  2023, \mn@doi [Astronomische Nachrichten]
  {https://doi.org/10.1002/asna.20220131}, 344, e220131

\bibitem[\protect\citeauthoryear{{Varniere} \& {Vincent}}{{Varniere} \&
  {Vincent}}{2017}]{VV17}
{Varniere} P.,  {Vincent} F.~H.,  2017, \mn@doi [\apj]
  {10.3847/1538-4357/834/2/188}, \href
  {https://ui.adsabs.harvard.edu/abs/2017ApJ...834..188V} {834, 188}

\bibitem[\protect\citeauthoryear{{Varniere}, {Tagger}  \&
  {Rodriguez}}{{Varniere} et~al.}{2011}]{VTR11}
{Varniere} P.,  {Tagger} M.,   {Rodriguez} J.,  2011, \mn@doi [\aap]
  {10.1051/0004-6361/201015028}, \href
  {http://cdsads.u-strasbg.fr/abs/2011A%26A...525A..87V} {525, A87}

\bibitem[\protect\citeauthoryear{{Varniere}, {Casse}  \& {Vincent}}{{Varniere}
  et~al.}{2018}]{VCH18XMM}
{Varniere} P.,  {Casse} F.,   {Vincent} F.~H.,  2018, in Proceedings of the
  XMM-Newton 2018 Science Workshop, TIME-DOMAIN ASTRONOMY: A HIGH ENERGY VIEW
  ESAC, MADRID, SPAIN, 13 - 15 JUNE 2018.

\bibitem[\protect\citeauthoryear{{Varniere}, {Vincent}  \& {Casse}}{{Varniere}
  et~al.}{2020}]{V20}
{Varniere} P.,  {Vincent} F.~H.,   {Casse} F.,  2020, \mn@doi [\aap]
  {10.1051/0004-6361/202037816}, \href
  {https://ui.adsabs.harvard.edu/abs/2020A&A...638A..33V} {638, A33}

\bibitem[\protect\citeauthoryear{Varniere, Casse  \& Dodu}{Varniere
  et~al.}{2024}]{VCDA24}
Varniere P.,  Casse F.,   Dodu F.,  2024, A\&A, in prep

\bibitem[\protect\citeauthoryear{{Vincent}, {Paumard}, {Gourgoulhon}  \&
  {Perrin}}{{Vincent} et~al.}{2011}]{Vin11}
{Vincent} F.~H.,  {Paumard} T.,  {Gourgoulhon} E.,   {Perrin} G.,  2011,
  \mn@doi [Classical and Quantum Gravity] {10.1088/0264-9381/28/22/225011},
  \href {http://adsabs.harvard.edu/abs/2011CQGra..28v5011V} {28, 225011}

\bibitem[\protect\citeauthoryear{{Westernacher-Schneider}, {Zrake}, {MacFadyen}
   \& {Haiman}}{{Westernacher-Schneider}
  et~al.}{2022}]{WesternacherSchneider22}
{Westernacher-Schneider} J.~R.,  {Zrake} J.,  {MacFadyen} A.,   {Haiman} Z.,
  2022, \mn@doi [\prd] {10.1103/PhysRevD.106.103010}, \href
  {https://ui.adsabs.harvard.edu/abs/2022PhRvD.106j3010W} {106, 103010}

\makeatother
\end{thebibliography}

\end{document}